\documentstyle[epsf,amssymb,aps]{revtex}

\newcommand{\mafigura}[4]{
  \begin{figure}[hbtp]
    \begin{center}
      \epsfxsize=#1 \leavevmode \epsffile{#2}
    \end{center}
    \caption{#3}
    \label{#4}
  \end{figure} }

\begin{document}
\author{L. B. Leinson}
\address{Institute of Terrestrial Magnetism, Ionosphere and 
Radio Wave Propagation\\
RAS, 142092 Troitsk, Moscow Region, Russia\\
e-mail : leinson@charley.izmiran.rssi.ru\\
Phone : (7 095) 3340912 \\
Fax: (7 095) 3340124}
\author{A. P\'{e}rez}
\address{Departamento de F\'{\i }sica Te\'{o}rica, Universidad de Valencia \\
46100 Burjassot (Valencia) Spain\\
e-mail : Armando.Perez@uv.es\\
Phone : (34 96) 3864593 \\
Fax:  (34 96) 3642345}
\title{Direct URCA process in neutron stars with strong magnetic fields. }
\maketitle

\begin{abstract}
We calculate the emissivity for the direct URCA process in strongly
magnetized, degenerate matter in neutron stars, under $\beta $-equilibrium.
We show that, if the magnetic field is large enough for protons and
electrons to be confined to the ground Landau levels, the field-free
threshold condition on proton concentration no longer holds, and direct URCA
reactions are open for an arbitrary proton concentration. Direct URCA
process leads to an early phase of fast neutron star cooling. This
circumstance allows us to constrain the initial magnetic field inside
observed pulsars{\bf .}
\end{abstract}

\vspace{1 cm}


\section{\protect\bigskip Introduction}

It is now well established that pulsars are neutron stars with a very strong
magnetic field. The surface magnetic field manifests itself by synchrotron
radiation from the pulsar magnetosphere and its magnitude for young neutron
stars has been estimated \cite{Wolfer} to be as large as $10^{12}-10^{14}$\
G. The internal magnetic field may not manifest at the surface and therefore
its strength is controversial but, because of the high conductivity of the
core, the value of the magnetic field is expected to be further increased .
Considering the flux throughout the star as a simply trapped primordial
flux, one obtains an internal magnetic field intensity of the order $10^{18}$%
\ G, even if one neglects a possible twisting of the field, an effect that
would intensify the field additionally by several orders of magnitude. The
time duration of the large-scale magnetic field decay via the mechanism of 
ambipolar diffusion  might be about a few
decades or a hundred year \cite{Haensel}. Such a simple hypothesis meets,
however, a reasonable criticism, because nascent neutron stars are
convective. Some works have been devoted to develop an alternative to the
origin of the internal magnetic field \cite{Thompson}, based on the
convective dynamo mechanism, which gives rise to a magnetic field of the
order $10^{14}-10^{15}$\ G inside neutron stars. However, in spite of the
fact that we do not know yet any appropriate mechanism to produce more
intense magnetic fields, the scalar virial theorem \cite{Lai} allows the
field magnitude to be as large as $10^{18}$\ G.  Magnetic field of this
magnitude, and even more intense, have been recently discussed \cite{Ternov}%
, \cite{Daicic}, \cite{Bocquet}, \cite{Serot}, \cite{Vshivtsev} as possibly
existing inside some neutron stars and other compact astrophysical objects.
To constrain the magnetic field inside neutron stars, it would be helpful to
search for any observable consequences of a superstrong magnetic field. For
example, one can observe a rapid motion of the neutron star under the recoil
action, due to anisotropic neutrino emission induced by an intense magnetic
field \cite{Chugay}, \cite{Dorofeev} or one can analyze the effect of a
magnetic field on the neutron star mass \cite{Vshivtsev}.

Moreover, if superstrong magnetic fields exist in the interior of young
neutron stars, new processes would be open, and standard processes would be
also modified by the magnetic field, thus changing the evolution of those
stars. Studying the influence of these modifications and confrontation with
observational data can be used to put some bounds on the initial magnetic
field. The aim of this paper is to consider one of these modifications, and
to motivate further research on this field. 

As we will show, a crucial mechanism when considering the cooling of neutron
stars with superstrong magnetic fields are the direct URCA reactions 
\begin{equation}
p+e^{-}\stackrel{B}{\rightarrow }n+\nu   \label{pe}
\end{equation}
\begin{equation}
n\stackrel{B}{\rightarrow }p+e^{-}+\overline{\nu }\;  \label{n}
\end{equation}
where the symbol $B$ on the top of the arrow means that the reactions rates
have to be calculated for particle states in the magnetic field.

We will consider degenerate, nonrelativistic nuclear matter and
ultrarelativistic degenerate electrons under $\beta $-equilibrium. Matter is
assumed to be totally transparent for neutrinos and antineutrinos. We also
assume that at zero temperature the neutron star is $\beta $-stable, but at
non-zero temperature $T\ll \varepsilon _i^F$ ($i=n,p,e$) reactions (\ref{pe}%
) and (\ref{n}) proceed near the Fermi energies $\varepsilon _i^F$  of the
participating particles. In the field-free case, for reactions $%
p+e^{-}\rightarrow n+\nu $ ; $n\rightarrow p+e^{-}+\overline{\nu }$ to take
place, momentum conservation near the Fermi surfaces requires the following
inequality among the Fermi momenta of the proton ($p_F$), the electron ($k_F$%
) and neutron ($q_F$): 
\begin{equation}
p_F+k_F\geq q_F  \label{1}
\end{equation}
Together with the charge neutrality condition, the above inequality leads to
the threshold for proton concentration $Y_p\equiv n_p/\left( n_p+n_n\right)
\geq 1/9$, where $n_p$ and $n_n$ are the number densities of protons and
neutrons, respectively. This means that, in the field-free case, direct URCA
reactions are strongly suppressed by Pauli blocking in the neutron-rich
nuclear matter (assuming matter is formed only by protons, neutrons and
electrons).

In magnetized media, transversal momenta of charged particles are defined
only within an accuracy $\Delta p_{\perp }\sim \sqrt{eB}$. We show in the
following that strong magnetic fields $\sqrt{eB}\sim q_F$ break down
condition Eq. (\ref{1}) so that direct URCA processes are allowed in the
degenerate n-p-e system under $\beta $-equilibrium for an arbitrary proton
concentration. We examine reactions (\ref{pe}) and (\ref{n}) in a constant
magnetic field large enough so that \footnote{%
Here $e$ is the absolute value of electron charge. For electrons having a
Fermi momentum of $50$ Mev one obtains $B>2\times 10^{17}$ G. We set $\hbar
=c=k_{B}=1$.} $B>k_F^2/\left( 2e\right) $ . In this case the Fermi energy of
degenerate electrons is smaller than the energy of the first excited Landau
level $\sqrt{m^2+2eB}$ . In other words, at zero temperature, all electrons
occupy only the ground Landau band and are polarized in the direction
opposite to the magnetic field . The number density of electrons reads 
\begin{equation}
n_e=\frac{eB}{4\pi ^2}\int_{-k_F}^{k_F}dk_3=\frac{eB}{2\pi ^2}k_F  \label{ne}
\end{equation}
where $k_F$ is the Fermi momentum of degenerate electrons. Following charge
neutrality $n_p=n_e$, one obtains the proton Fermi momentum $p_F$ to be
equal to $k_F$. Therefore, if $B>k_F^2/\left( 2e\right) $, then degenerate
protons are also confined to the ground Landau state and are totally
polarized in the direction parallel to the external magnetic field. We
choose our coordinate system in such a way that ${\bf B}=\left( 0,0,B\right) 
$. As will be obtained through the next sections, under the conditions
mentioned above the direct URCA process (\ref{pe}) and (\ref{n}) gives raise
to a large neutrino luminosity during the first 10 yr. of the strongly
magnetized neutron star. Such large luminosity would lead to an anomalously
cold neutron star, a fact that can be confronted with observational data and
eventually used to put some bounds on the initial magnetic field.

\section{Calculation of the emissivity}

The emissivity for reactions (\ref{pe}) and (\ref{n}) can be computed using
standard charged-current $\beta $-decay theory. The matrix element for the
V-A interaction reads : 
\begin{equation}
M_\sigma =\frac G{\sqrt{2}}\int d^3r\overline{\Psi }_{n\sigma }\left( {\bf r}%
\right) \gamma _\mu \left( 1-g_A\gamma _5\right) \Psi _p\left( {\bf r}%
\right) \overline{\Psi }_\nu \left( {\bf r}\right) \gamma ^\mu \left(
1-\gamma _5\right) \Psi _e\left( {\bf r}\right)   \label{Me}
\end{equation}
where $\sigma =+$ ($\sigma =-$) corresponds to neutrons polarized along (in
the opposite direction to) the magnetic field, $\Psi _\nu \left( {\bf r}%
\right) $ and $\Psi _e$ represent the neutrino and electron fields ,
respectively, while $\Psi _p$ and $\Psi _n$ stand for the proton and
neutron; $G=G_F\cos \theta _C$ with $\theta _C$ being the Cabibbo angle, and 
$g_A\simeq 1.261$ is the axial-vector coupling constant. By using the
asymmetric (Landau) gauge ${\bf A=}(0,Bx,0)$ , the four-dimensional
polarized wave functions can be expressed in terms of stationary states in
the normalization volume $V=L_1L_2L_3$. The electron wave function,
corresponding to the ground Landau band with energy $\varepsilon _0=\sqrt{%
m^2+k_3^2}$ reads 
\begin{equation}
\Psi _e\left( {\bf r}\right) =\frac 1{\sqrt{2\varepsilon _0}\sqrt{L_2L_3}}%
\exp (-i\varepsilon _0t)\exp \left[ i\left( k_3z+k_2y\right) \right] \varphi
_0(\xi )u_e  \label{evf}
\end{equation}
The function $\varphi _0(\xi )$ is the eigenfunction of the one-dimensional
harmonic oscillator, normalized with respect to $\xi =\sqrt{eB}\left(
x+k_2/\left( eB\right) \right) $ 
\begin{equation}
\varphi _0(\xi )=%
{eB \overwithdelims() \pi }
^{1/4}{}\exp (-eB\xi ^2/2)\,  \label{3}
\end{equation}
and 
\begin{equation}
u_e=\frac 1{\sqrt{\varepsilon _0+m}}\left( 
\begin{tabular}{l}
$0$ \\ 
$\varepsilon _0+m$ \\ 
$0$ \\ 
$-k_3$%
\end{tabular}
\right)   \label{ue}
\end{equation}
with $k_2,k_3$ the electron momenta in the $y$ and $z$ directions,
respectively. In order to take into account strong interactions in nuclear
matter, we consider nonrelativistic protons, with an effective mass $M_p^{*}$%
, moving in the self-consistent uniform nuclear potential $U_p$. Then the
proton wave function reads 
\begin{equation}
\Psi _p\left( {\bf r}\right) =\frac 1{\sqrt{L_2L_3}}\exp (-iE_0t)\exp \left[
i\left( p_3z+p_2y\right) \right] \varphi _0\left( \zeta \right) \left( 
\begin{array}{c}
W_{+} \\ 
0
\end{array}
\right)   \label{4}
\end{equation}
where $p_2,p_3$ are proton momenta in the $y$ and $z$ directions,
respectively, and $\zeta =\sqrt{eB}\left( x-p_2/\left( eB\right) \right) $.
Polarized protons are given by the non-relativistic spin-up bispinor $W_{+}$
normalized by the condition $W_{+}^{\dagger }W_{+}=1$. By introducing the
anomalous magnetic moment of the proton, its energy reads \cite{Itzykson} $%
E_0\simeq \widetilde{M}_p+\varepsilon _p+U_p$ where $\varepsilon
_p=p_3^2/\left( 2\widetilde{M}_p\right) $ and 
\begin{equation}
\widetilde{M}_p=M_p^{*}-\frac e{2M_p^{*}}\left( g_p-1\right) B\,  \label{7}
\end{equation}
with the proton's Lande factor $g_p=2.79$. In a similar way, we consider
non-relativistic neutrons, of effective mass $M_n^{*}$, moving in the
self-consistent uniform nuclear potential $U_n$. One can describe neutron
polarization by non-relativistic bispinors $W_\sigma $ ($\sigma =\pm $).
They are normalized by the condition $W_\sigma ^{\dagger }W_\sigma =1$. Then
the neutron wave function reads 
\begin{equation}
\Psi _{n\sigma }\left( {\bf r}\right) =\frac 1{\sqrt{L_2L_3L_3}}\exp
(-iE_{q\sigma }t)\exp \left[ i\left( q_3z+q_2y+q_1x\right) \right] \left( 
\begin{array}{c}
W_\sigma  \\ 
0
\end{array}
\right)   \label{10}
\end{equation}
Here the energy of a neutron with spin polarization $\sigma $ is $E_{q\sigma
}=M_n^{*}+\varepsilon _{q\sigma }+U_n$ with 
\begin{equation}
\varepsilon _{q\sigma }=\frac{q^2}{2M_n^{*}}-\sigma \mu _nB  \label{11}
\end{equation}
which incorporates the neutron interaction with the external magnetic field
due to its anomalous magnetic moment $\mu _n=g_ne/\left( 2M_p\right) $ with $%
g_n=-1.91$ . In accordance to Eq.(\ref{11}), neutrons with different
polarizations $\sigma =\pm $ have different momenta $q_{F\pm }$ at the Fermi
surface $\varepsilon _{q\pm }=\varepsilon _n^F$. Taking into account 
\begin{equation}
\varepsilon _n^F=\frac{q_{F\pm }^2}{2M_n^{*}}\pm \mu _nB  \label{EFn}
\end{equation}
one finds $q_{F-}^2=q_{F+}^2+4M_n^{*}\mu _nB$. Explicit evaluation of the
matrix elements $M_\sigma $ in (\ref{Me}) yields 
\begin{equation}
\left. 
\begin{array}{c}
\left| M_{+}\right| ^2 \\ 
\left| M_{-}\right| ^2
\end{array}
\right\} =\left. 
\begin{array}{c}
2\left( 1+g_A^2\right) \left( \omega +\kappa _3\right)  \\ 
8g_A^2\left( \omega -\kappa _3\right) 
\end{array}
\right\} G^2\left( \varepsilon _0+k_3\right) \exp \left( -\frac 12\frac{%
q_1^2+\left( p_2+k_2\right) ^2}{eB}\right)   \label{Mpsq}
\end{equation}
The neutrino momentum is ${\bf \kappa }=(\kappa _1,\kappa _2,\kappa _3)$ and
its energy is $\omega =\left| {\bf \kappa }\right| $. We have neglected the
neutrino momentum $\kappa =\omega \sim T$, when compared to neutron momenta $%
q_{F\pm }$. The neutrino emissivity reads :

\begin{eqnarray}
\epsilon _{\nu } &=&\frac{1}{V^{2}L_{2}^{2}L_{3}^{2}}%
\int_{-eBL_{1}/2}^{eBL_{1}/2}\frac{L_{2}dk_{2}}{2\pi }\int \frac{L_{3}dk_{3}%
}{2\pi }\int_{-eBL_{1}/2}^{eBL_{1}/2}\frac{L_{2}dp_{2}}{2\pi }\int \frac{%
L_{3}dp_{3}}{2\pi }\int \frac{Vd^{3}q}{\left( 2\pi \right) ^{3}}\int \frac{%
Vd^{3}\kappa }{\left( 2\pi \right) ^{3}}  \label{Emis} \\
&&\times \frac{\omega }{2\omega 2\varepsilon _{0}}2\pi \left[ 
\begin{array}{c}
\left| M_{+}\right| ^{2}\delta \left( E_{q+}+\omega -E_{0}-\varepsilon
_{0}\right) \left( 1-f_{n+}\right) f_{e}f_{p} \\ 
+\left| M_{-}\right| ^{2}\delta \left( E_{q-}+\omega -E_{0}-\varepsilon
_{0}\right) \left( 1-f_{n-}\right) f_{e}f_{p}
\end{array}
\right]  \nonumber \\
&&\times 2\pi \delta \left( q_{3}+\kappa _{3}-p_{3}-k_{3}\right) 2\pi \delta
\left( q_{2}+\kappa _{2}-p_{2}-k_{2}\right)  \nonumber
\end{eqnarray}
Here $\left( 1-f_{n\sigma }\right) f_{e}f_{p}$ are statistical factors
corresponding to Fermi-Dirac distributions of particles with chemical
potentials $\psi _{n}\simeq M_{n}^{\ast }+U_{n}+\varepsilon _{n}^{F}$, $\psi
_{e}\simeq \varepsilon _{e}^{F}$ and $\psi _{p}\simeq \widetilde{M}%
_{p}+\varepsilon _{p}^{F}+U_{p}$. Fermi energy of neutrons is defined in (%
\ref{EFn}). For electrons and protons they are given by : 
\begin{equation}
\varepsilon _{e}^{F}=\sqrt{m^{2}+k_{F}^{2}}\hspace{1in}\varepsilon _{p}^{F}=%
\frac{p_{F}^{2}}{2\widetilde{M}_{p}}  \label{Fen}
\end{equation}
With these definitions one has 
\begin{equation}
f_{n\pm }=\frac{1}{\exp \left[ \left( \varepsilon _{q\pm }-\varepsilon
_{n}^{F}\right) /T\right] +1}\;  \label{fn}
\end{equation}
\begin{equation}
f_{e}=\frac{1}{\exp \left[ \left( \varepsilon _{0}-\varepsilon
_{e}^{F}\right) /T\right] +1}  \label{fe}
\end{equation}
\begin{equation}
f_{p}=\frac{1}{\exp \left[ \left( \varepsilon _{p}-\varepsilon
_{p}^{F}\right) /T\right] +1}  \label{fp}
\end{equation}
Since the integrand in Eq. (\ref{Emis}) depends on $p_{2}$ and $k_{2}$ only
through the combination $p_{2}+k_{2}$, we can introduce a new variable $%
k_{2}+p_{2}\rightarrow p_{2}$ and perform integration over $k_{2}$. We can
also neglect the small neutrino momentum in the momentum conservation $%
\delta $-functions. Thus, we obtain 
\begin{eqnarray}
\epsilon _{\nu } &=&\frac{G^{2}eB}{\left( 2\pi \right) ^{7}}\int
dk_{3}dp_{2}dp_{3}d^{3}qd^{3}\kappa \,\delta \left( q_{3}-p_{3}-k_{3}\right)
\delta \left( q_{2}-p_{2}\right) \frac{\left( \varepsilon _{0}+k_{3}\right) 
}{\varepsilon _{0}}\exp \left( -\frac{q_{1}^{2}+p_{2}^{2}}{2eB}\right) 
\nonumber \\
&&\times \left[ 
\begin{array}{c}
\frac{1}{2}\left( 1+g_{A}^{2}\right) \left( \omega +\kappa _{3}\right)
\delta \left( M_{n}^{\ast }-\widetilde{M}_{p}+U_{n}-U_{p}+\varepsilon
_{q+}+\omega -\varepsilon _{p}-\varepsilon _{0}\right) \left(
1-f_{n+}\right) f_{e}f_{p} \\ 
+2g_{A}^{2}\left( \omega -\kappa _{3}\right) \delta \left( M_{n}^{\ast }-%
\widetilde{M}_{p}+U_{n}-U_{p}+\varepsilon _{q-}+\omega -\varepsilon
_{p}-\varepsilon _{0}\right) \left( 1-f_{n-}\right) f_{e}f_{p}
\end{array}
\right]  \label{15}
\end{eqnarray}
The integrals can be performed in a straightforward way by using the
following identity 
\begin{eqnarray}
&&f_{e}\left( \varepsilon _{0}\right) f_{p}\left( \varepsilon _{p}\right)
\left( 1-f_{n}\left( \varepsilon _{n\pm }\right) \right) \delta \left(
M_{n}^{\ast }-\widetilde{M}_{p}+U_{n}-U_{p}+\varepsilon _{q\pm }+\omega
-\varepsilon _{p}-\varepsilon _{0}\right)  \nonumber \\
&=&\int_{-\infty }^{\infty }d\varepsilon _{3}\int_{-\infty }^{\infty
}d\varepsilon _{2}\int_{-\infty }^{\infty }d\varepsilon _{1}f_{e}\left(
\varepsilon _{1}\right) f_{p}\left( \varepsilon _{2}\right) \left(
1-f_{n}\left( \varepsilon _{3}\right) \right)  \nonumber \\
&&\times \delta \left( M_{n}^{\ast }-\widetilde{M}_{p}+U_{n}-U_{p}+%
\varepsilon _{3}+\omega -\varepsilon _{2}-\varepsilon _{1}\right) \delta
\left( \varepsilon _{3}-\varepsilon _{n\pm }\right) \delta \left(
\varepsilon _{2}-\varepsilon _{p}\right) \delta \left( \varepsilon
_{1}-\varepsilon _{0}\right)  \label{Ident}
\end{eqnarray}
Then, due to Pauli blocking, the main contribution to the integrals comes
from $\varepsilon _{1}\approx \varepsilon _{e}^{F}$, $\varepsilon
_{2}\approx \varepsilon _{p}^{F}$ and $\varepsilon _{3}\approx \varepsilon
_{n}^{F}$ . By this reason, one can substitute $\delta \left( \varepsilon
_{n}^{F}-\varepsilon _{n\pm }\right) \delta \left( \varepsilon
_{p}^{F}-\varepsilon _{p}\right) \delta \left( \varepsilon
_{e}^{F}-\varepsilon _{0}\right) $instead of $\delta \left( \varepsilon
_{3}-\varepsilon _{n\pm }\right) \delta \left( \varepsilon _{2}-\varepsilon
_{p}\right) \delta \left( \varepsilon _{1}-\varepsilon _{0}\right) $.
Considering electrons as ultrarelativistic, we obtain that $\left(
\varepsilon _{0}+k_{3}\right) $ is not small only when for $k_{3}\geq 0$.

Under the $\beta $-equilibrium assumption, one has the following
relationship : 
\begin{equation}
f_e\left( \varepsilon _0\right) f_p\left( \varepsilon _p\right) \left(
1-f_n\left( \varepsilon _{n\pm }\right) \right) =\left( 1-f_e\left(
\varepsilon _0\right) \right) \left( 1-f_p\left( \varepsilon _p\right)
\right) f_n\left( \varepsilon _{n\pm }\right)   \label{fequil}
\end{equation}
Therefore, the neutron decay process (\ref{n}) gives the same energy loss
rate as process (\ref{pe}), although giving final antineutrinos. We finally
arrive to the following total neutrino plus antineutrino emissivity: 
\begin{eqnarray}
\epsilon  &=&\epsilon _0\frac{M_n^{*}}{p_F}\frac{\widetilde{M}_p}{M_p}\frac B%
{B_0}\times   \nonumber \\
&&[\left( 1+g_A^2\right) \Theta \left( q_{F+}^2-4p_F^2\right) \exp \left( -%
\frac{q_{F+}^2-4p_F^2}{2eB}\right) +4\left( 1+g_A^2\right) \Theta \left(
q_{F+}^2\right) \exp \left( -\frac{q_{F+}^2}{2eB}\right)   \nonumber \\
&&+2g_A^2\Theta \left( q_{F-}^2-4p_F^2\right) \exp \left( -\frac{%
q_{F-}^2-4p_F^2}{2eB}\right) +8g_A^2\Theta \left( q_{F-}^2\right) \exp
\left( -\frac{q_{F-}^2}{2eB}\right) ]  \label{finnumeric}
\end{eqnarray}
Where the charge neutrality condition $k_F=p_F$ was used and $\epsilon _0=2%
\frac{457\pi }{40320}G^2M_p^3T^6=1.04\times 10^{27}\,T_9^6\,\text{(ergs/cm}^3%
\text{s)}$, with $T_9$ the temperature in units of $10^9K$ and $%
B_0=M_p^2/e\simeq 1.5\times 10^{20}G$. Since both electrons and protons are
at the ground Landau level, the third-component momentum conservation reads
either $q_3=k_F+p_F$, when proton and electron go in the same direction,
along the magnetic field, or $q_3=k_F-p_F$ when the initial particles move
in opposite directions . Since $k_F=p_F$ and $q_{F\pm }^2>q_3^2$, four
possibilities are open : $q_{F+}^2>\left( 2p_F\right) ^2$, $q_{F-}^2>\left(
2p_F\right) ^2$ , $q_{F+}^2>0$ and $q_{F-}^2>0$. This is the origin of the
four terms in the latter equation, which are multiplied by their
corresponding step functions, where \thinspace $\Theta \left( x\right) $
gives $+1$ if the argument exceeds $0$, and is $0$ otherwise. As we will
show later, opening or closing of these channels will give rise to jumps on
the emissivity.

\section{Results}

In order to estimate numerically the emissivity of the direct URCA process
in a strongly magnetized neutron star, we consider nuclear matter and
electrons in $\beta $-equilibrium within a nonrelativistic Hartree approach
in the linear $\sigma $-$\omega $-$\rho $ model, as it was made
(relativistically) in \cite{Serot}, but with taking into account the
anomalous magnetic moments of proton and neutron in the energies for these
particles. Then the interaction energies for protons and neutrons are given
by 
\begin{eqnarray}
U_p &=&\left( \frac{g_\omega }{m_\omega }\right) ^2n_b+\frac 14\left( \frac{%
g_\rho }{m_\rho }\right) ^2\rho _3\,  \label{Up} \\
U_n &=&\left( \frac{g_\omega }{m_\omega }\right) ^2n_b-\frac 14\left( \frac{%
g_\rho }{m_\rho }\right) ^2\rho _3  \label{Un}
\end{eqnarray}
where $n_b=n_p+n_n$ is the total number density of baryons and $\rho
_3=n_p-n_n$. Also, 
\begin{eqnarray}
n_p &=&n_e=\frac{M_p^2}{2\pi ^2}\frac B{B_0}p_F\,  \label{np} \\
n_n &=&\frac 1{6\pi ^2}\left( q_{F+}^3+q_{F-}^3\right)   \label{nn}
\end{eqnarray}
The parameters for the coupling constants and mesons masses are taken from 
\cite{Horovitz} to be $g_\omega ^2\left( M/m_\omega \right) ^2=273.87$, and $%
g_\rho ^2\left( M/m_\rho \right) ^2=97$. In this model, mass $M=939MeV$ is
the same for proton and neutron. 

To make our estimates for the emissivity, we solved for variables $q_{F+}\,$%
and $p_F$ the $\beta $-equilibrium condition 
\begin{equation}
\psi _p+\psi _e=\psi _n  \label{beta}
\end{equation}
with the help of equations (\ref{Up}-\ref{nn}) and $%
q_{F-}^2=q_{F+}^2+4M_n^{*}\mu _nB$, for a given baryon density $n_b$. In
order to be fully consistent with incorporating magnetic moments of proton
and neutron, one would also like to find the nucleon effective mass $M^{*}$
from this modified model. However, this would add some (perhaps unessential)
calculation problems. Our purpose here is to point out some qualitative new
results and motivate more realistic computations. For this reason , we have
chosen for our estimates the values of the effective mass that are given by 
\cite{Serot}. More precisely, we have taken $M^{*}=0.6M$ for $n_b$ equal to
nuclear saturation density, which is $n_0=0.15fm^{-3}$ in this model, and $%
M^{*}=0.3M$ for $n_b=2n_0$.

In Fig. 1 we have plotted the emissivity of the direct URCA process,
calculated from Eq. (\ref{finnumeric}) as a function of the magnetic field
intensity (in units of $B/B_{0}$) for two values of the baryon density : $%
n_{b}=n_{0}$ and $n_{b}=2n_{0}$. We have chosen, as a representative value
for the temperature, $T=10^{9}K$. As one can see from this figure, if the
magnetic field is larger than $B\simeq 7\times 10^{17}G$ , the $\nu 
\overline{\nu }$ emissivity from the neutron star is many orders of
magnitude larger than that in the standard cooling scenario. Another feature
which is readily observed is that, as magnetic field increases, jumps on the
emissivity can appear. The first jump in the curve $n_{b}=n_{0}$ is due to $%
q_{F-}^{2}-4p_{F}^{2}$ becoming negative when $B/B_{0}\gtrsim 0.014$. The
second, much smaller one, at $B/B_{0}\sim 0.026$ , corresponds to $%
q_{F-}^{2} $ becoming zero. For the curve $n_{b}=2n_{0}$ we observe a
different situation : $q_{F-}^{2}-4p_{F}^{2}$ is always negative, but an
abrupt increase appears due to the fact that $q_{F+}^{2}-4p_{F}^{2}$ is
first negative and, as magnetic field increases, it becomes positive at $%
B/B_{0}\sim 0.03$.

Due to the transitions discussed above, it becomes difficult to find an
overall analytic formula to fit the emissivity dependence upon the magnetic
field intensity, although it is possible to find a fitting formula which
describes the gross tendency of $\epsilon $ (without the details described
above) for a particular value of the baryon density. We have found, when $%
n_{b}=n_{0}$ that the emissivity can be approximated by : 
\begin{equation}
\epsilon \simeq 2.40\times 10^{27}b\left( 2.855-4.255b\right) \left(
23.05e^{-0.037/b}+5.763e^{-0.02/b}\right) T_{9}^{6}erg\,cm^{-3}s^{-1}
\label{enb}
\end{equation}
where $b=B/B_{0}$ .

We now discuss the implications of the large emissivity produced by the URCA
reactions in the presence of a strong magnetic field. Of course, in order to
make detailed calculations, one needs to incorporate this emissivity into an
elaborated neutron star cooling code. Aside from this, it is clear that
other processes will also be modified in presence of the strong magnetic
field (and maybe new processes will be possible). However, such
considerations are out of the scope of this paper, as we discussed in the
introduction. Here, we will only present some simple estimates.

As it is evident from Fig. 1, the emissivity depends very strongly on the
magnetic field intensity. One expects that an initial strong $B$ will decay
within a time $t_{B}$ . Hence, even if the luminosity due to the URCA
process is very high at the beginning, it will become very small for $t>>$ $%
t_{B}$. According to \cite{Haensel}, the characteristic decaying time for a
superstrong magnetic field is $t_{B}\sim 100$ yr. In order to incorporate
this effect, we have taken, as the time variation of $B$ : 
\begin{equation}
B(t)=B_{initial}e^{-t/t_{B}}
\end{equation}

Let us also assume that the thermal energy $U_{th}$ of matter can be
approximated by that of a degenerate neutron gas, then per unit volume one
has : 
\begin{equation}
u_{th}=\frac{U_{th}}{V}=\frac{1}{2}AT^{2}
\end{equation}
Since a neutron interacts with the external magnetic field only due to its
anomalous magnetic moment, the thermal energy of a degenerate neutron gas
will only slightly change under the action of the magnetic field. Neglecting
this difference we assume\ $A=\left( \frac{\pi }{3}\right)
^{2/3}n_{b}^{1/3}M_{n}^{\ast }$.

One has to solve the equation 
\begin{equation}
\frac{du_{th}}{dt}=-\epsilon
\end{equation}

which can be rewritten as 
\begin{equation}
\frac{dT}{dt}=-\frac{\epsilon }{(du_{th}/dT)}  \label{diffT}
\end{equation}

For our estimate, we will consider a fixed value of the density, with $%
n_b=n_0$ . Then $u_{th}\sim 6.0\times 10^{28}T_9^2\;erg\,cm^{-3}$ . We have
also chosen an initial value of the magnetic field $B_{initial}=7\times
10^{17}G$ and the temperature $T(0)=10^9K$. For the emissivity, we use the
approximate formula Eq. (\ref{enb}). Equation (\ref{diffT}) can then be
solved numerically. The result is plotted in Fig. 2, where the vertical axes
represents $T_9$ and the abscissa gives the time (in years). As seen from
this figure, the temperature drops very fast at the beginning due to the
URCA emission. However, as $B$ goes to zero, the emission will decrease and
the temperature stabilizes to a value which, in this case, is of the order
of $10^8K$ . In Fig. 3 we have plotted the URCA luminosity that arises from
the above simplified model, with an assumed neutron star radius $R=10^6cm$.
Here, one observes the large luminosity at early times and its rapid
decrease due to the temperature drop. There is an ulterior secular decrease
of the luminosity as a consequence of the magnetic field decay, so that $%
L_{URCA}$ will become negligeable for $t>>t_B$ . Standard emission processes
will then dominate the neutron star cooling.

\section{Conclusions}

In this work we have considered the direct URCA process in degenerate
nuclear matter under $\beta $-equilibrium in a strong magnetic field. These
reactions are strongly suppressed by Pauli blocking in the field-free case,
but we showed that, in a magnetic field $B\gtrsim k_F^2/\left( 2e\right) $,
direct URCA reactions are open for an arbitrary proton concentration, and
will lead to a phase of fast neutron star cooling. A simple estimate shows
that, after a rapid cooling epoch (around 10 yr. or so), the neutron star
temperature becomes one order of magnitude lower than in the field-free
case, a result which is in discrepancy with observations of pulsar data 
\footnote{%
For a review on pulsar data, see for example{\bf \ }\cite{Becker}.}. One
then can conclude that neutron stars with characteristic cooling times of
the order $10^5-10^6$\ years, as in the standard scenario, could not have an
initial magnetic field comparable or larger than the critical value $%
B_c\gtrsim k_F^2/\left( 2e\right) $, at least \ in the whole volume of the
central core. On the other hand, if such stars exist, one has to search for
anomalously cold neutron stars. \ In order to investigate the implications
of this process, we have considered a very simple model. To simplify
calculations we also restricted ourselves to the above values of the
magnetic field. This approach makes possible to treat all protons and
electrons as occupying only the ground Landau bands at zero temperature.
However, as one can see from Fig. 1, the direct URCA process can
substantially contribute to neutron star cooling, even if the magnetic field
is smaller than the value considered above. For the URCA processes to be
effective one needs the electron momentum uncertainty $\Delta k_{\perp }\sim 
\sqrt{eB}$\ to be of the order of the transversal momentum itself $k_{\perp
}\sim \sqrt{2neB}$. This condition is still valid when electrons occupy a
few Landau bands with $n\sim 1$.\  We also have considered matter densities
not much larger than saturation densities. For these densities, nucleons can
be treated as non-relativistic particles. More realistic calculation imply
additional complications. If one intends to go to densities larger than the
saturation density, then relativistic effects for nucleons should be taken
into account and the many-level approximation should be used in order to
calculate the energy loss via the direct URCA process. The standard
processes should be also modified with taking into account a superstrong
magnetic field. By making such a detailed simulation one could put more
stringent bounds on the initial intensity of the magnetic field.

\acknowledgments
This work has been partially supported by Spanish DGICYT Grant PB94-0973 and
CICYT AEN96-1718. L.B. L. would like to thank for partial support of this
work by RFFR Grant 97-02-16501 and INTAS Grant 96-0659. We are grateful to 
J. A. Miralles for his comments.

\newpage 

\mafigura{7cm}{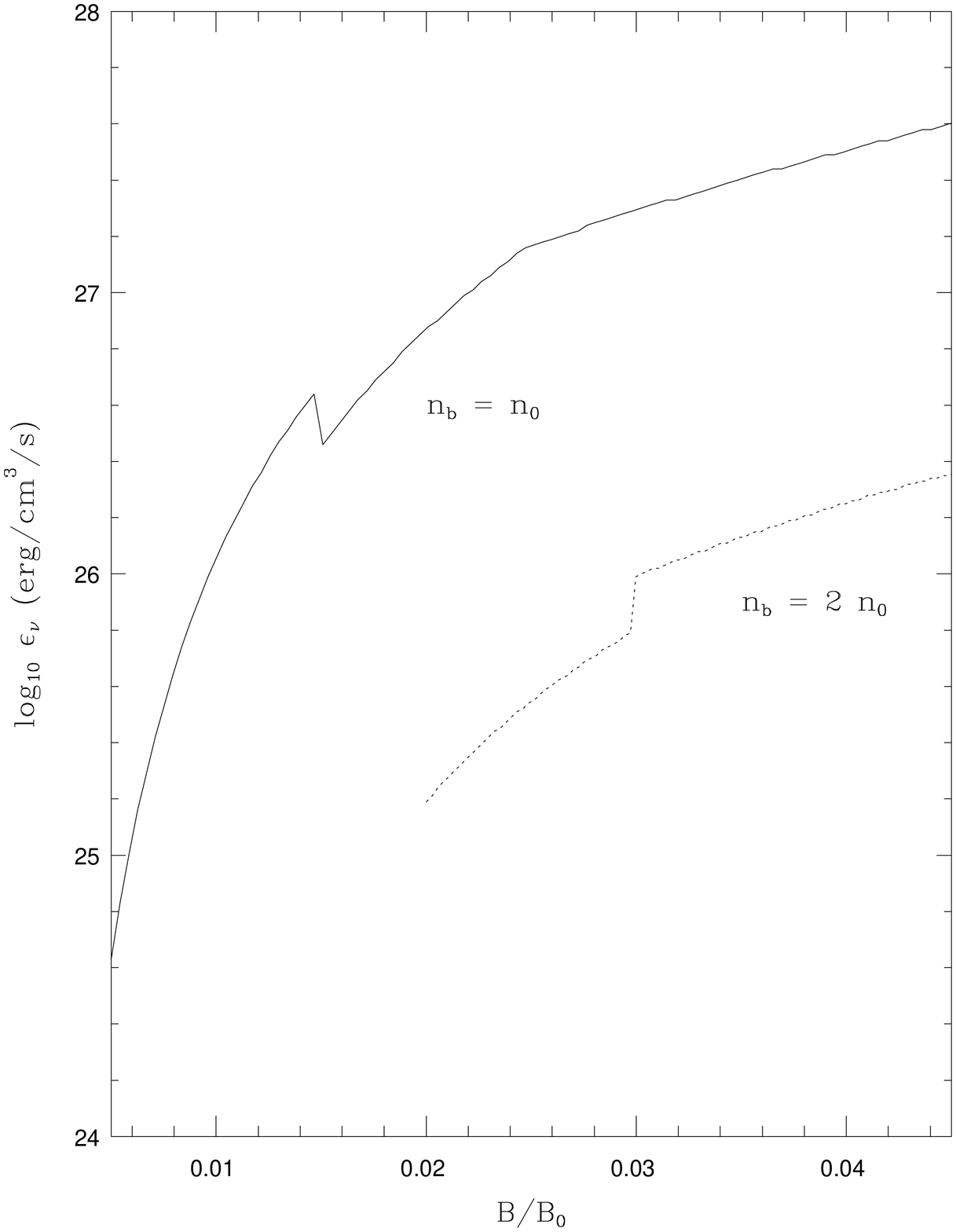}{Neutrino emissivity of the direct URCA process as a
function of the magnetic field intensity $B$ ($B$ in units of the proton
critical field $B_0$), for two values of the baryon density and for a
temperature $T=10^9\ K$.}{Fig. 1}


\mafigura{7cm}{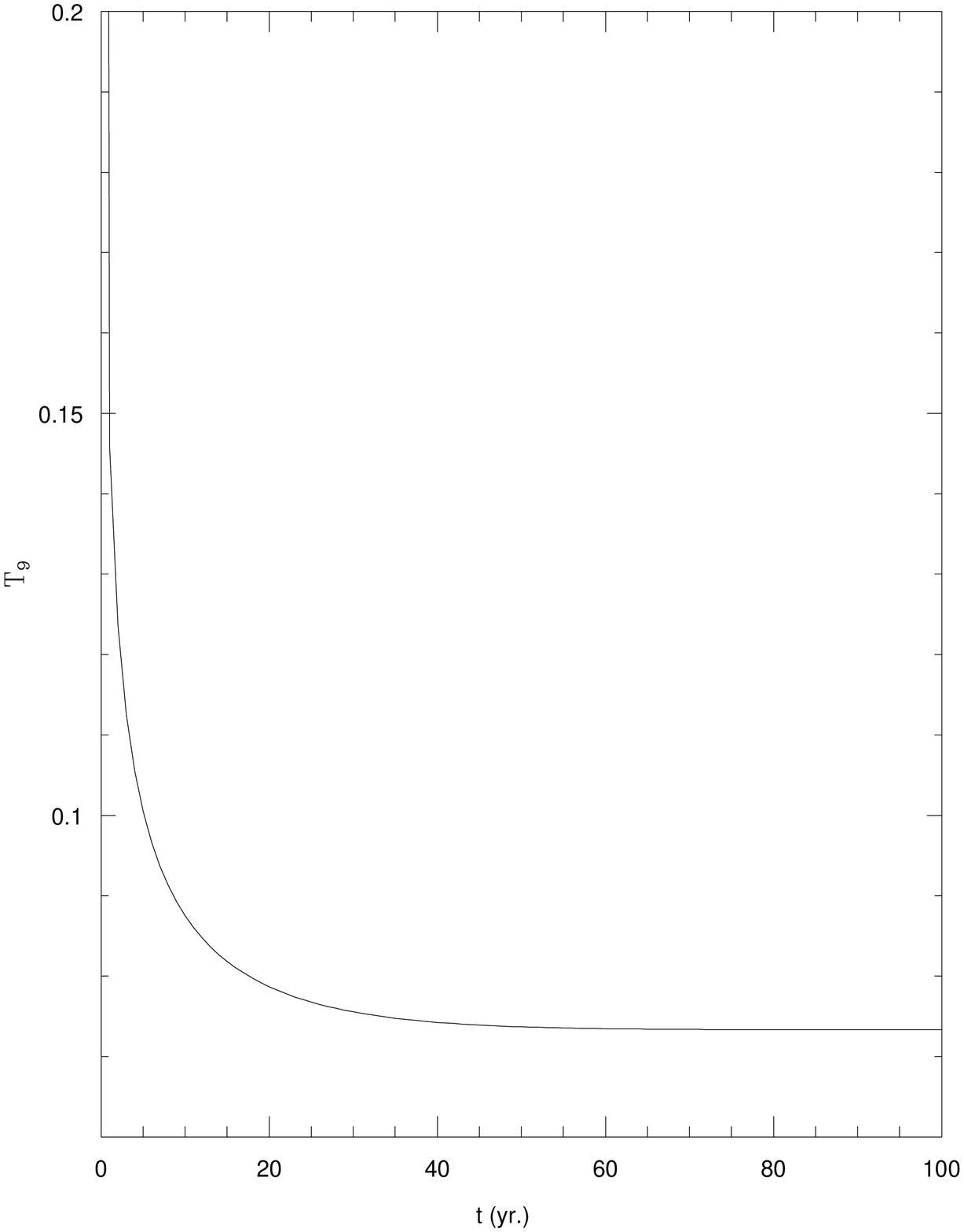}{Temperature evolution, in units of $10^9K$ for the 
simple model discussed on the text.}{Fig. 2}


\mafigura{7cm}{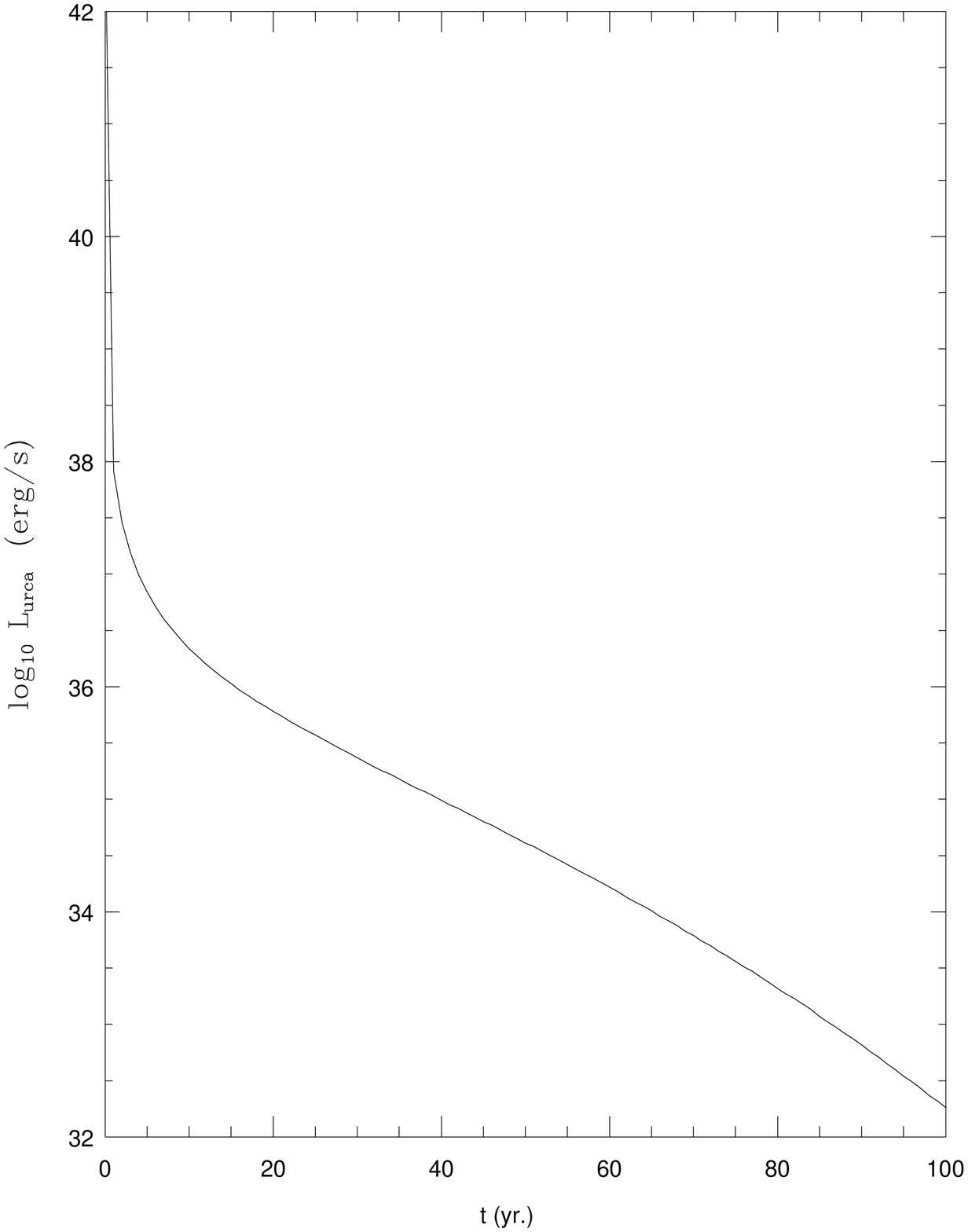}{Luminosity of the URCA process (in $erg/s$) as 
a function of time, for the same model, in logarithmic scale.}{Fig. 3}

\end{document}